\documentclass[12pt]{article}
\usepackage{cite}
\usepackage{amsmath,amssymb,amsfonts}
\usepackage{graphicx,color} 
\usepackage{textcomp}
\usepackage{xcolor}
\usepackage{hyperref}
\hypersetup{hidelinks}
\usepackage{comment}
\usepackage{algorithm}
\usepackage{algorithmic}
\usepackage[top=1in,bottom=1in,left=1in,right=1in]{geometry}

\DeclareMathOperator*{\argmin}{arg\,min}
\DeclareMathOperator*{\argmax}{arg\,max}

\usepackage{bm}
\usepackage{bbm}
\newcommand{\dB}{[\text{dB}]}

\newcommand{\rev}[1]{\textcolor{black}{#1}}

\begin{document}

\title{Image detection using combinatorial auction}

\author{Simon Anuk, Tamir Bendory, and Amichai Painsky}

\maketitle

\begin{abstract}
    This paper studies \rev{the optimal solution of} the classical problem of detecting the location of multiple image occurrences in a two-dimensional, noisy measurement.
   Assuming the image occurrences do not overlap, we formulate this task as a constrained 
    maximum likelihood optimization problem. We show that the maximum likelihood estimator is equivalent to an instance of the winner determination problem from the field of combinatorial auction and that the solution can be obtained by searching over a binary tree. 
    We then design a pruning mechanism that significantly accelerates the runtime of the search. 
    We demonstrate on simulations and electron microscopy data sets that the proposed algorithm provides accurate detection in challenging regimes of high noise levels and densely packed image occurrences. 
\end{abstract}

\section{Introduction}
This paper studies the problem of accurately detecting multiple occurrences of a template image $s\in \mathbb{R}^{W \times W}$ located in a noisy measurement $y$ \rev{(the observed, acquired, data)}. 
Specifically, let $y \in \mathbb{R}^{N\times M}$  be a measurement of the form:
\begin{equation}\label{Pixels_in_pic}
    {y[n,m]} = \sum_{k=1}^{K}s\left[n-c^{(k)}_{n},m-c^{(k)}_{m}\right] + w[n,m],
\end{equation}
where ${\mathbf{c}^{(k)}}= (c^{(k)}_{n}, c^{(k)}_{m})$ is the upper-left corner coordinate of the $k$-th image occurrence. We model the noise term  $w[n,m]$ as an i.i.d.\ Gaussian noise with mean zero and variance $\sigma^2$.
Our goal is to estimate the unknown image locations ${\mathbf{c}^{(1)}},\ldots,{\mathbf{c}^{(K)}}$ from \rev{the observation} $y$. 
While $K$---the number of image occurrences---is an unknown parameter, it is instructive to first assume that $K$ is known; we later omit this assumption.

\rev{The problem of detecting multiple image occurrences in noisy data appears in various image processing applications, including fluorescence microscopy~\cite{geisler2007resolution, egner2007fluorescence}, astronomy~\cite{brutti2005spike}, anomaly detection~\cite{ehret2019image}, neuroimaging~\cite{worsley1996unified,genovese2002thresholding,taylor2007detecting}, and electron microscopy~\cite{eldar2020klt,heimowitz2018apple,bepler2019positive}. A detailed discussion on these applications is provided in~\cite{cheng2017multiple}.}
\rev{However, although many algorithms were developed over the years (see, e.g.~\cite{ehret2019image}) in most cases there is no known statistically optimal method for detecting the images.}

\begin{figure}
  \centering
  \begin{minipage}[b]{0.48\columnwidth}
    \centering
    \includegraphics[width=.8\linewidth]{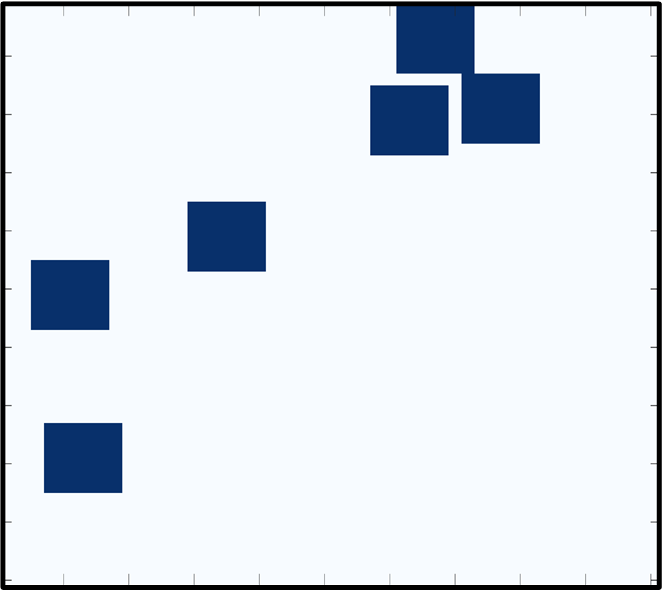}
    \label{fig:clean measurement}
  \end{minipage}
  \hfill
  \begin{minipage}[b]{0.48\columnwidth}
    \centering
    \includegraphics[width=.8\linewidth]{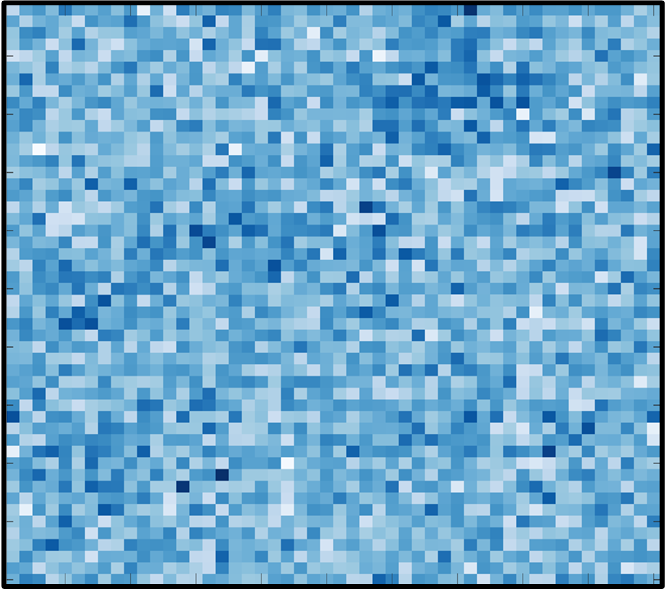}
    \label{fig:noisy measurement}
  \end{minipage}
  \caption{The left panel displays a noiseless measurement with six image occurrences: the images on the right side are densely packed, while the images on the left side are separated by more than $2W$ pixels. The right panel shows a noisy version of the same measurement, with SNR=$-30\dB$. Our goal is to accurately estimate the location of the six images from the noisy measurement.}
  \label{fig:An-example-of-a-clean-measurement-and-a-noisy-measurement}
\end{figure}

\rev{A particularly important motivation for this paper is single-particle cryo-electron microscopy (cryo-EM)---a leading technology for determining the three-dimensional structure of biological molecules~\cite{bendory2020single,singer2020computational,frank2006three}.}
\rev{In a cryo-EM experiment, multiple samples of biological molecules are frozen in a thin layer of ice. The samples might be densely packed, but they do not overlap.
One of the first stages in the computational pipeline of cryo-EM involves locating images within a large and highly noisy measurement~\cite{eldar2020klt,heimowitz2018apple,bepler2019positive}.} 
\rev{Motivated by cryo-EM,} we assume that the image occurrences do not overlap, but otherwise may be arbitrarily spread in the measurement. 
Namely, each image occurrence is separated by at least $W$ pixels from any other image occurrence, in at least one dimension:
\begin{equation}
\begin{aligned}
    \label{eq:Dense Separation Condition}  
    \max\left(\left| c^{(k)}_{n} - c^{(\ell)}_{n}\right|,\left|c^{(k)}_{m} - c^{(\ell)}_{m}\right|\right) \geq W,  \quad  \forall k \neq \ell. 
\end{aligned}
\end{equation}
 We refer to~\eqref{eq:Dense Separation Condition} as the \emph{separation condition}.
Figure~\ref{fig:An-example-of-a-clean-measurement-and-a-noisy-measurement} illustrates an example of a measurement that contains six image occurrences. 
The three image occurrences on the right side are densely packed (but satisfy the separation condition~\eqref{eq:Dense Separation Condition}), while the image occurrences on the left are separated by more than  $2W$ pixels (twice the separation condition~\eqref{eq:Dense Separation Condition}); the distinction between these two scenarios plays a key role in this paper.

Assuming the template image $s$ and the number of image occurrences $K$ are known, 
the maximum likelihood estimator (MLE), for the sought locations, is given by \begin{equation}
    \label{eq:Maximum likelihood}
    \begin{aligned}
    \argmin_{\hat{\mathbf{c}}^{(1)},\ldots,\hat{\mathbf{c}}^{(K)}}\left\| y - \sum_{k=1}^{K}s\left[n-\hat{c}^{(k)}_{n},m-\hat{c}^{(k)}_{m}\right] \right\|^{2}_{2},
    \end{aligned}
\end{equation}
where we denote by $\hat{\mathbf{c}}^{(1)},\ldots,\hat{\mathbf{c}}^{(K)}$ the estimates of ${\mathbf{c}}^{(1)},\ldots $ ${\mathbf{c}}^{(K)}$, respectively. Taking the separation condition~\eqref{eq:Dense Separation Condition} into account, the MLE results in a constrained optimization problem:
\begin{equation} 
    \label{eq:constrainted_optmization}
    \begin{aligned}        &\argmax_{\hat{\mathbf{c}}^{(1)},\ldots,\hat{\mathbf{c}}^{(K)}}\sum_{k=1}^{K}\sum_{n=1}^{N}\sum_{m=1}^{M}y[n,m]s\left[n-\hat{c}^{(k)}_{n},m-\hat{c}^{(k)}_{m}\right]\\
        & \ \textrm{s.t.} \quad \max\left(\left|\hat{c}^{(k)}_{n} - \hat{c}^{(\ell)}_{n}\right|,\left|\hat{c}^{(k)}_{m} - \hat{c}^{(\ell)}_{m}\right|\right) \geq W,  \quad \forall k \neq \ell.
    \end{aligned}
\end{equation}
Efficiently computing the optimal solution to this optimization problem is the \rev{prime goal} of this work.

A popular and natural heuristic for solving~\eqref{eq:constrainted_optmization} is as follows~\cite{heimowitz2018apple,eldar2020klt,prasad2018detection,fukunishi2018improvements,roth2023detecting}.
First, we correlate the template image $s$ with the measurement~$y$. We set $\hat{\mathbf{c}}^{(1)}$ as the location which maximizes the correlation. The next estimator, $\hat{\mathbf{c}}^{(2)}$, is chosen as the next maximum of the correlation that satisfies the separation condition~\eqref{eq:Dense Separation Condition} with respect to ${\hat{\mathbf{c}}^{(1)}}$. The same approach is applied sequentially for estimating the rest of the locations $\hat{\mathbf{c}}^{(3)},\ldots,\hat{\mathbf{c}}^{(K)}$. We refer to this algorithm as the \textit{greedy algorithm}, and it can be thought of as a variation of the well-known template matching algorithm that takes the separation condition~\eqref{eq:Dense Separation Condition} into account. 
This algorithm is highly efficient since the correlations can be calculated using the FFT algorithm~\cite{rapuano2007introduction}.

\begin{figure}
  \centering
  \begin{minipage}[b]{0.48\columnwidth}
    \centering
    \includegraphics[width=.8\linewidth]{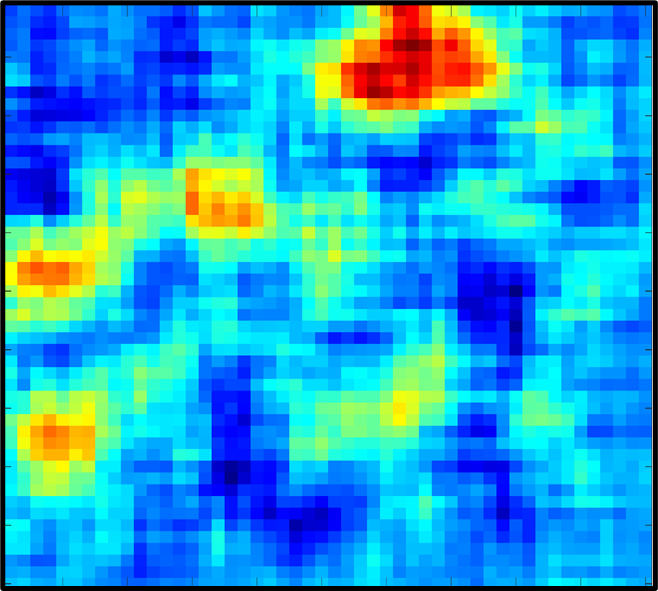}
    \label{fig:corr_heatmap}
  \end{minipage}
  \hfill
  \begin{minipage}[b]{0.48\columnwidth}
    \centering
    \includegraphics[width=.8\linewidth]{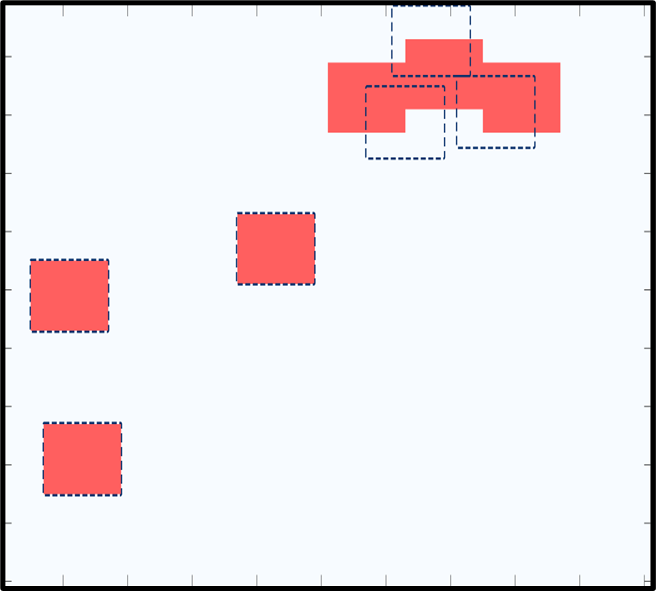}
    \label{fig:greedy_output}
  \end{minipage}
  \caption{On the left, the correlation between the noisy measurement~$y$, as depicted in Figure ~\ref{fig:An-example-of-a-clean-measurement-and-a-noisy-measurement}, and the template image $s$. On the right, the dashed lines display the noiseless measurement, as depicted in Figure~\ref{fig:An-example-of-a-clean-measurement-and-a-noisy-measurement}. The red images are the output of the greedy algorithm, which fails to detect the location of the images on the right when the image occurrences are densely packed.}
  \label{fig:figures}
\end{figure}

Figure~\ref{fig:figures} illustrates the correlation between the noisy measurement $y$ of Figure~\ref{fig:An-example-of-a-clean-measurement-and-a-noisy-measurement} and the template image $s$, \rev{and the} output of the greedy algorithm. 
Evidently, the greedy algorithm successfully detects the locations of the well-separated image occurrences on the left end of the measurement  (when the images are separated by at least $2W$ pixels, twice the separation condition of~\eqref{eq:Dense Separation Condition}) but completely fails when the image occurrences are densely packed (right end).
The reason is that correlation doubles the support of the image. Thus, the greedy algorithm is likely to consider two close image occurrences as one, especially in the presence of high levels of noise.
In this paper, we design an algorithm that 
detects the image occurrences accurately for any measurement that satisfies the separation condition~\eqref{eq:Dense Separation Condition}---even for measurements with highly dense image occurrences---by computing the constrained MLE~\eqref{eq:constrainted_optmization}.

While many papers designed object detection algorithms, as far as we know, none of them is guaranteed to achieve the MLE~\cite{ehret2019image}, as we propose in this paper. For example, a CNN-based technique was designed in~\cite{ren2015faster}; this method works well but has no theoretical guarantees and requires a large data set for training. Other papers focused on accelerating the template matching algorithm~\cite{mahmood2011correlation}. The MLE for the one-dimensional version of~\eqref{Pixels_in_pic} was considered in~\cite{roth2023detecting} based on dynamic programming. 
A multiple hypothesis approach was suggested in~\cite{schwartzman2011multiple,cheng2017multiple,eldar2024object}, and the statistical properties of the problem were analyzed in~\cite{dadon2024detection}.

In Section~\ref{sec:ca}, we reformulate the constrained MLE problem~\eqref{eq:constrainted_optmization} 
as a Winner Determination Problem (WDP)---an optimization problem from the field of combinatorial auction. 
Section~\ref{sec:algorithm} shows how the problem can be solved exactly by binary tree search. While scanning the entire tree is intractable, we design an efficient pruning method to significantly accelerate the search, while guaranteeing to find the optimal solution. This section outlines the technical details of the proposed algorithm. 
Section~\ref{sec:Numerical_Experiments} presents results on simulated data, and Section~\ref{sec:Cryo} shows results on experimental electron microscopy datasets. Section~\ref{sec:conclusions} concludes the paper.

\section{Image detection as a combinatorial auction problem} \label{sec:ca}

This section begins by introducing the classical combinatorial auction problem and the WDP. Then, we show how the constrained optimization problem~\eqref{eq:constrainted_optmization} can be reformulated as an optimization problem, \rev{which is} highly similar to the classical WDP. 
This formulation leads to an efficient algorithm, 
whose details are introduced in Section~\ref{sec:algorithm}.  

\subsection{Classical combinatorial auction}
Consider an auction of a variety of different goods. Unlike a classical auction, where goods are sold individually, a bidder can place a bid on a bundle of goods in a combinatorial auction. The auction manager receives multiple price offers for different
bundles of goods. The goal of the auction manager is to maximize his revenue by allocating the goods to the highest submitted bids. Naturally, the same good may not be allocated to multiple bidders. This means that if two bidders place bids over bundles with overlapping goods, only one of the bids \rev{can} win.

Let $\mathcal{G} = \{\gamma_{1},\ldots,\gamma_{A}\}$ be a set of given goods, and let $\mathcal{B} = \{\beta_{1},\ldots,\beta_{B}\}$  be a set of bids. Each bid $\beta_i$ includes a bundle of goods $g(\beta_i)\subseteq \mathcal{G}$, which the bidder seeks to acquire. For this bundle, the bidder of $\beta_i$ proposes a price, $p(\beta_i)$. Let $x_i$ be an indicator variable that corresponds to the winning bids. That is, we set $x_i=1$ if $\beta_i$ is a winning bid and $x_i=0$ otherwise. An allocation, $\bm{\pi}$, is a set of winning bids that do not overlap. Specifically, $\bm{\pi}\subseteq \mathcal{B}$
is a subset of the bids, where $\forall \beta_2 \neq \beta_1 \in \bm{\pi}, g(\beta_1)\cap g(\beta_2) = \varnothing$. Therefore, an allocation cannot include two bids that include the same good. 

Using this notation, the combinatorial auction problem is given by: 
\begin{equation} \label{eq:combinatorial}
\begin{aligned}
&\max_{x_1,\ldots,x_{B}} && \quad \sum_{i=1}^{B} x_{i}p(\beta_{i})\\
& \; \; \;\textrm{s.t.} &&\sum_{i\vert \gamma \in g(\beta_{i})}^{}x_{i} \leq 1, \quad &&\forall \gamma \in \mathcal{G},\\
&&& \; \; \; \; x_{i} \in \{0, 1\}, &&\forall i.
\end{aligned}
\end{equation}
The first constraint ensures that each good can be allocated once at most, while the second constraint guarantees that each bid can be either included or excluded from the chosen allocation.
We define the \textit{optimal allocation} as the allocation that maximizes the revenue of the auction manager.
The problem of determining the optimal allocation is known as the WDP~\cite{leyton2003resource}.
It is known that the WDP is an \textit{NP}-hard problem and 
cannot be approximated, within any constant, by a polynomial-time algorithm~\cite{rothkopf1998computationally,sandholm2002algorithm}.

\subsection{Image detection as a winner determination problem}

We now cast the constrained optimization problem~\eqref{eq:constrainted_optmization} as a variation of the WDP. We associate every pixel in the measurement with a single good and let $\mathcal{B}$ be a set of bids.  Accepting $\beta_{ij} \in \mathcal{B}$ implies that our estimator includes an image occurrence whose upper left corner is located at the $(i, j)$-th pixel. The goods requested by the bid, 
$g(\beta_{ij})$, are a $W\times W$ set of pixels, \rev{corresponding} to a possible image occurrence in the measurement. The offered price of the bid, $p(\beta_{ij})$, is the correlation between the measurement and the template image, namely,
\begin{equation}
\label{Correlation Z}
   p(\beta_{ij}) = \sum_{n=1}^{N}\sum_{m=1}^{M}y[n,m]s[n-i,m-j]. 
\end{equation}
The revenue of the auctioneer is the sum of offered prices by all accepted bids.  Since we are estimating the upper left corner coordinates of images in the measurement, we require that the estimated coordinates will satisfy $\hat{c}^{(k)}_{n} \leq N-W+1$ and $\hat{c}^{(k)}_{m} \leq M-W+1$, so we do not exceed the measurement's dimension. We denote $\tilde{N} = N-W+1$ and $\tilde{M} = M-W+1$.

Let us define $\mathbf{X}\in\{0,1\}^{\tilde{N}\times \tilde{M}}$ as a matrix of indicators, in which
$x_{ij} \in \mathbf{X}$ is equal to $1$ if the bid $\beta_{ij}$ is included in the allocation, and $0$ otherwise. 
To enforce the separation constraint \eqref{eq:Dense Separation Condition}, we require that 
\begin{equation} 
\label{2-D separation condition}
\begin{aligned} \mathbbm{1}_{_{W}}^T\Tilde{\mathbf{X}}_{ij}\mathbbm{1}_{_{W}} \in \{0,1\}, \quad \forall i,j,
\end{aligned}
\end{equation}
where $\mathbbm{1}_W$ is an all-ones $W\times1$ column vector and $\Tilde{\mathbf{X}}_{ij}$ is a $W\times W$ sub-matrix of $\mathbf{X}$, whose upper left entry is the $(i,j)$ entry in the matrix $\mathbf{X}$. 
Using this notation, and assuming $K$ is known, the constrained likelihood problem~\eqref{eq:constrainted_optmization} can be written as:
\begin{equation} \label{eq:WDP K bids}
\begin{aligned}
&\max_{x_1,\ldots,x_{\tilde{N}\tilde{M}}} && \sum_{i=1}^{\tilde{N}}\sum_{j=1}^{\tilde{M}}x_{ij}p(\beta_{ij})\\
&\text{s.t.} &&\mathbbm{1}_{_{\tilde{N}}}^T\mathbf{X}\mathbbm{1}_{_{\tilde{M}}}= K,\\
&&& \mathbbm{1}_{_{W}}^T\Tilde{\mathbf{X}}_{ij}\mathbbm{1}_{_{W}} \leq 1, && \forall i,j,\\
&&& x_{ij} \in \{0, 1\}, && \forall i,j,
\end{aligned}
\end{equation}
where the last two conditions are, together, equivalent to~\eqref{2-D separation condition}. 
Therefore, solving~\eqref{eq:WDP K bids} is equivalent to~\eqref{eq:constrainted_optmization}, where the non-zero entries of the optimal \rev{$\mathbf{X}$} correspond to the MLE of the locations in~\eqref{eq:constrainted_optmization}.

We underscore that the optimization problem~\eqref{eq:WDP K bids} is slightly different from the classical combinatorial auction~\eqref{eq:combinatorial}. In the classical setup, 
the auctioneer aims to include as many bids as possible in the optimal allocation to maximize the revenue, whereas in our model the optimal allocation must include exactly $K$ bids. Furthermore, in our model, $p(\beta_{ij})$ can be negative.  
Thus, we cannot use existing algorithms that solve the classical WDP~\eqref{eq:combinatorial}, such as the one proposed in~\cite{leyton2003resource}, and we design a new algorithm tailored for our constrained setup.
Importantly, the designed algorithm must be computationally efficient:  solving~\eqref{eq:WDP K bids} in a brute-force manner requires searching over 
an exponential number of bids, rendering this approach intractable.  For example, if $N,M = 30$ and $K = 5$, then the number of possible allocations is $\sim 10^{12}$.
The next section introduces a pruning technique to solve~\eqref{eq:WDP K bids}, \rev{which dramatically reduces} the number of explored allocations.

\section{WDP for image detection}
\label{sec:algorithm}
To delineate the proposed algorithm, we introduce further notation. 
Two  bids, $\beta_i, \beta_j \in \mathcal{B}$, are considered to be conflicting if $g(\beta_i)\cap g(\beta_j) \neq  \varnothing$.  We extend the functions $g(\cdot)$ and $p(\cdot)$ to apply to allocations: $g(\bm{\pi}) = \bigcup_{\beta\in\bm{\pi}}g(\beta)$ and $p(\bm{\pi}) = \sum_{\beta\in\bm{\pi}}p(\beta)$  A partial allocation, ${\bm{\pi}_k}$, is an allocation in which the number of bids is $k < K$. 
A bid $\beta_i$ is considered to be conflicting with a partial allocation ${\bm{\pi}_k}$ if $g(\beta_i) \cap g({\bm{\pi}_k}) \neq  \varnothing$. We denote the set of bids that conflict with a partial allocation ${\bm{\pi}_k}$ as $C_{\bm{\pi}_k}$. Hence, given a partial allocation ${\bm{\pi}_k}$, the set of allowed bids, from which a bid can be added to ${\bm{\pi}_k}$, is $\overline{C}_{{\bm{\pi}_k}} = \mathcal{B}\setminus C_{{\bm{\pi}_k}}$. 
In addition, we define $\tilde{\bm{\pi}}_\text{opt}$ as the current optimal allocation that was found by our algorithm so far.

\subsection{Binary tree construction}
The algorithm begins with forming a binary tree data structure. Binary trees, widely used in search algorithms in computer science~\cite{bentley1975multidimensional}, are suitable for examining every possible allocation through a systematic navigation to determine the optimal allocation.

\begin{figure}[h]   
	\centering \includegraphics[width=.8\columnwidth]{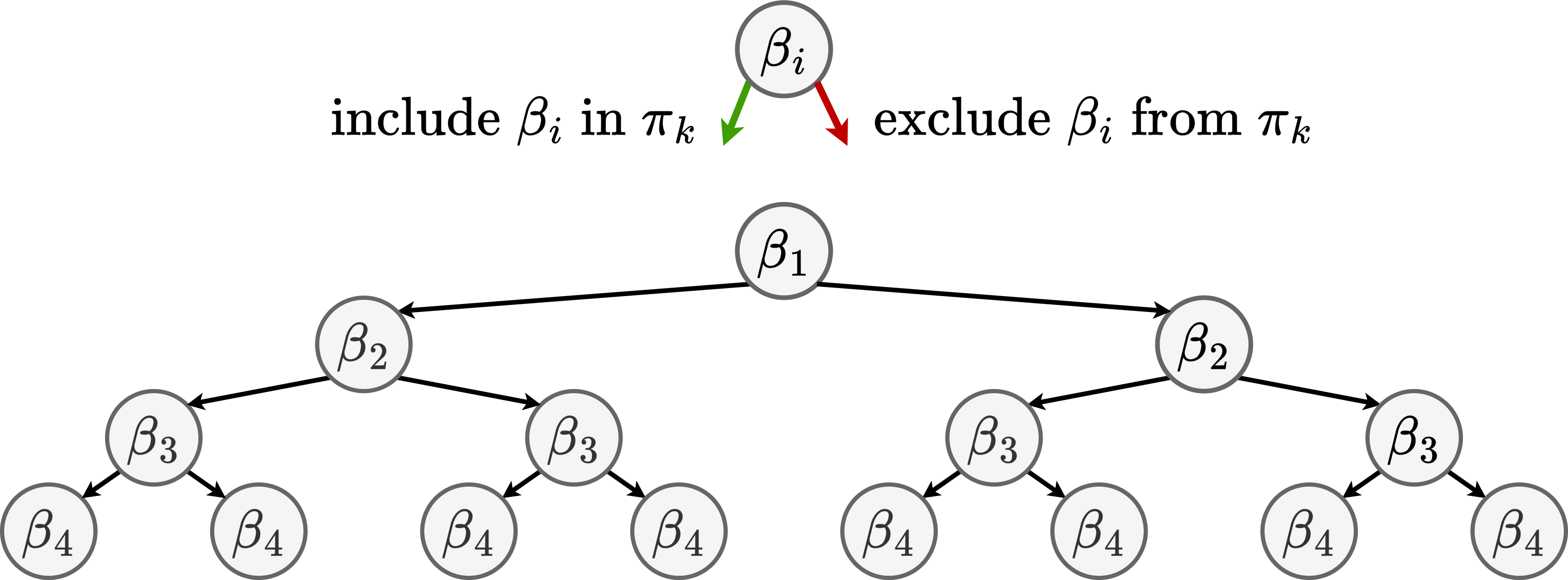}
    \caption{An example of a binary tree, corresponding to a measurement with four elements.}
    \label{fig:4 Bids Binary Tree}
\end{figure}

Given a set of bids, each node in the tree corresponds to a simple binary question: whether to include or exclude the bid in the allocation.
This results in a \textit{full binary tree} where, except for the leaf nodes, each node has two children corresponding to the same bid.
We introduce a simple example to illustrate the construction of the binary tree. Let us consider  $y\in \mathbb{R}^4$, a template image of a unit size, $s\in \mathbb{R}$, and  $K=2$. As explained in Section~\ref{sec:ca},  we cast the image detection problem as a variation of the WDP. In this example, every bid includes a bundle of a single good (since the template image is of a unit size).
We first set $\beta_1$ as the tree's root, and $\beta_4$ forms the tree's leaves. To explore an allocation that includes the current node, we navigate to the node's left child and add the current node to our partial allocation, $\bm{\pi}_{k}$. Conversely, navigating to the right child represents the alternative choice of excluding the current node from $\bm{\pi}_k$. Figure~\ref{fig:4 Bids Binary Tree} demonstrates the binary tree we obtain in this simple example. 

As previously mentioned, the binary tree enables us to conduct an exhaustive search across all possible allocations. Therefore, by using this search mechanism, the optimal allocation will inevitably be encountered during our search.

\subsection{Tree pruning}

Exhaustively searching through the tree leads necessarily to the optimal allocation, but is computationally infeasible even for problems of small dimensions. To overcome this issue, we use the algorithmic technique of \textit{branch-and-bound} and present a \textit{pruning} method to avoid exploring a large portion of the allocations in advance by eliminating sub-trees that cannot yield the optimal allocation.

\begin{figure}[h]   
	\centering \includegraphics[width=.8\columnwidth]{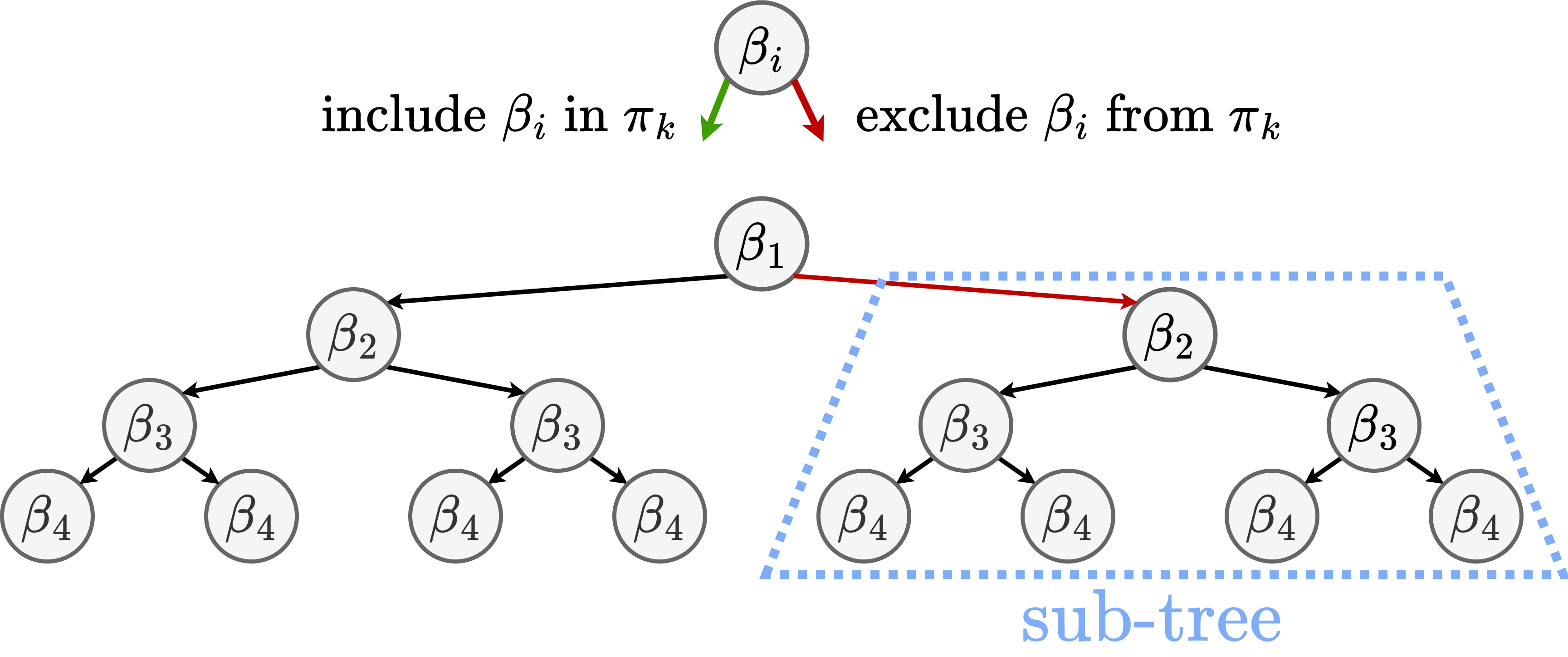}
    \caption{A sub-tree that corresponds to excluding $\beta_1$ from the partial allocation $\bm{\pi}_k$. If the maximal revenue that can be achieved by exploring this sub-tree is less than the revenue of our current optimal allocation, $\tilde{\bm{\pi}}_{\text{opt}}$, we prune it.}
    \label{fig:Sub-tree}
\end{figure}

We illustrate this method using the example of Figure~\ref{fig:4 Bids Binary Tree}. 
We refer to a smaller binary tree, spanned by a node other than the root, as a \textit{sub-tree}. Let us assume that our optimal allocation is $\bm{\pi}_\text{opt} = \{\beta_1,\beta_3\}$ and that we have already encountered our optimal allocation $\bm{\pi}_\text{opt}$ in our search. At this point, we are unaware that this is the optimal allocation, so we continue exploring the remaining possible allocations. Assume we have chosen to exclude $\beta_1$ from the empty partial allocation, $\bm{\pi}_0$. Following that, two bids are left to be found in the sub-tree spanned by~$\beta_2$. Figure~\ref{fig:Sub-tree} demonstrates this sub-tree. 
The maximal revenue achievable by exploring this sub-tree is bounded by the sum of the prices of the two bids with the highest revenue within it. Since this revenue is less than what is achieved by the optimal allocation (by the assumption that we already encountered the optimal allocation), we can avoid exploring the allocations in this sub-tree. Thus, in this example,  we reduce the computational burden by half, as we prune half of the binary tree.

We now formulate this example as a general pruning mechanism. 
Say we have reached a partial allocation~$\bm{\pi}_k$.
We sort the bids in $\overline{C}_{\bm{\pi}_k}$ (the set of allowed bids) in descending order according to their revenues.
We denote the $K-k$ maximal revenue bids in $\overline{C}_{\bm{\pi}_k}$ as $\overline{C}^{(K-k)}_{\bm{\pi}_k}$. Summing the revenues of the bids in $\overline{C}^{(K-k)}_{\bm{\pi}_k}$  bounds on the remaining revenue to be allocated in this sub-tree
\begin{equation}
h(\bm{\pi}_k) := \sum_{\beta\in\overline{C}^{(K-k)}_{\bm{\pi}_k}, }p(\beta).
\end{equation} 
 This function will enable us to determine if the optimal allocation cannot be achieved within the current sub-tree, and thus we can prune it.
Specifically, we prune the sub-tree if the sum of $p(\bm{\pi}_k)$ and $h(\bm{\pi}_k)$ is less than the revenue of our current optimal allocation, $p(\bm{\pi}_{\text{opt}})$, i.e.,
\begin{equation}
\label{Pruning Condition}
    p(\bm{\pi}_k) + h(\bm{\pi}_k) < p(\tilde{\bm{\pi}}_{\text{opt}}).
\end{equation}
Since we only prune sub-trees that do not lead to the optimal allocation, our proposed algorithm remains optimal. 
We note, however, that $h(\bm{\pi}_k)$ is not necessarily a tight upper bound as it ignores the constraint that the remaining $K-k$ bids must not conflict with each other.

\subsection{Sorting the bids} \label{sec:sorting}

Our algorithm examines each potential optimal allocation, and thus the optimality of the algorithm is unaffected by rearranging the order of the bids. We aim to reorder the bids in such a way that maximizes the efficiency of the pruning mechanism. 
Note that the earlier we secure a high revenue allocation $p(\tilde{\bm{\pi}}_{\text{opt}})$, the more the stopping criterion~\eqref{Pruning Condition} is met, resulting in fewer allocations to be examined.  
Hence, we begin the algorithm by sorting the bids by their revenue in descending order. The bid with the highest revenue is fixed as the tree's root, while the bid with the lowest revenue forms the tree's leaves. This strategy enhances the frequency in which the pruning condition is met, thereby boosting its efficiency and accelerating the algorithm's runtime. For example, in the experiments in Section~\ref{sec:Numerical_Experiments}, we observed that sorting the bids can lead to acceleration by a factor of around~10. 

\subsection{Algorithm description}
\label{Proposed Algorithm Subsection}

\begin{figure}[h]   
	\centering \includegraphics[width=.8\columnwidth]{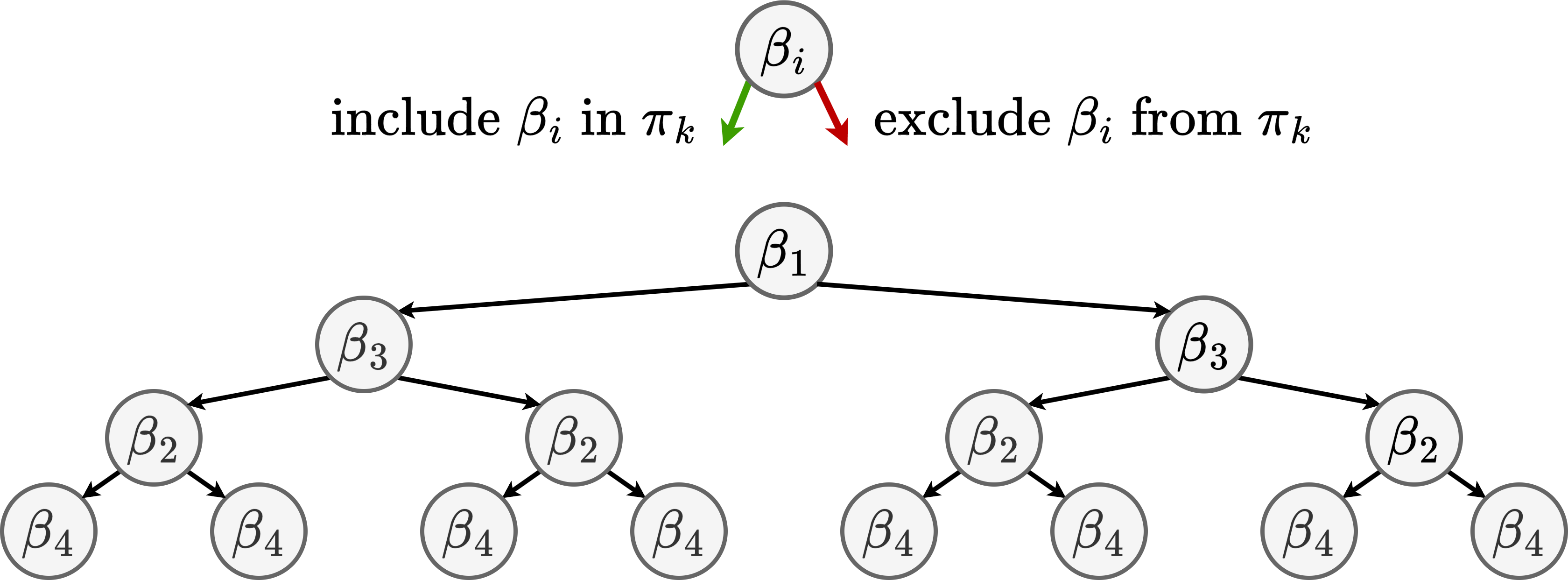}
    \caption{A sorted binary tree, corresponding to the tree that was presented in Figure~\ref{fig:4 Bids Binary Tree}.}
    \label{fig:Sorted Tree}
\end{figure}

We are now ready to describe our algorithm for detecting image occurrences in a noisy measurement. We will use the example above to explain the guiding principles.
We start by sorting the bids in descending order by their revenues, as described in Section~\ref{sec:sorting}.
An example of a sorted tree is presented in Figure~\ref{fig:Sorted Tree}.

Our algorithm is designed such that when we encounter a node, it is added to our partial allocation only if it does not conflict with the partial allocation and if the partial allocation itself does not satisfy the pruning condition~\eqref{Pruning Condition}.
We begin our search at the tree's root, $\beta_1$. As this is the first node we encounter, we add the node to our partial allocation and move to the left child, $\beta_3$.
At this point, $\beta_3$ is added to the partial allocation, since it does not conflict with our partial allocation nor does the partial allocation itself satisfy the pruning condition~\eqref{Pruning Condition}. Since this is the first allocation we attain, this is the current optimal allocation.
We note that the first allocation we explore is the output of the greedy algorithm. This means that the greedy algorithm is not necessarily optimal, since the optimal allocation may be achieved in a later stage. 
In our specific simple example, indeed the greedy algorithm attains the optimal solution (since the first two bids do not conflict), but in order to explain the full mechanism of the algorithm, we describe the remaining steps. 

We then remove the last bid we added, $\beta_3$, and continue examining allocations excluding that bid. To do so, we move to the right child of $\beta_3$. Upon reaching $\beta_2$, we check if our partial allocation $\bm{\pi}_1$  satisfies the pruning condition~\eqref{Pruning Condition}. 
Since the optimal allocation is $\bm{\pi}_\text{opt} = \{\beta_1,\beta_3\}$, the pruning condition is met, and hence we prune the sub-tree spanned by $\beta_2$ and return to the parent, $\beta_3$. Since we examined the sub-tree spanned by $\beta_3$, we next return to the parent, $\beta_1$, and remove $\beta_1$ from our partial allocation. 
This way, we examined all allocations that include $\beta_1$, and we move to the node's right child. This process continues until the algorithm explores the entire tree. Upon completion, it returns the optimal allocation.
The general algorithmic pipeline is described in Algorithm~\ref{alg}.

\begin{algorithm}[H]
    \caption{A WDP algorithm to solve~\eqref{eq:WDP K bids}} 
    \label{alg}
\begin{algorithmic}
  \STATE \textbf{Input:} {Set of bids $\mathcal{B}$ and number of image occurrences~$K$}
  \STATE \textbf{Output:} {Optimal allocation $\bm{\pi}_{\text{opt}}$}
\STATE Initialize: curr $\gets$ root,\ $\bm{\pi}_k = \{\}$,\ $\tilde{\bm{\pi}}_{\text{opt}} = \{\}$
\STATE Sort $\mathcal{B}$ by revenue and construct the binary tree
\WHILE{curr $\neq$ NULL}
    \IF{size$(\bm{\pi}_k) = K$}
         \IF{$p(\bm{\pi}_k) > p(\tilde{\bm{\pi}}_{\text{opt}})$}
            \STATE $\tilde{\bm{\pi}}_{\text{opt}} = \bm{\pi}_k$
         \ENDIF
         \STATE $\bm{\pi}_k \gets \bm{\pi}_k \setminus \text{curr.parent}$ 
         \STATE curr $\gets$ curr.parent
    \ELSE
        \IF{have not explored left sub-tree}
            \IF{$p(\bm{\pi}_k) + h(\bm{\pi}_k) < p(\tilde{\bm{\pi}}_{\text{opt}})$}   
                 \STATE prune left and right sub-trees.
            \ELSIF{curr $\in \overline{C}_{\bm{\pi}_k}$}
                \STATE $\bm{\pi}_k \gets \bm{\pi}_k\ \cup$ curr\ 
                \STATE curr $\gets$ curr.left
            \ENDIF
        \ELSIF{right sub-tree exists}
            \STATE curr $\gets$ curr.right \ 
        \ELSE
            \IF {curr $\in \bm{\pi}_k$}
                \STATE $\bm{\pi}_k \gets \bm{\pi}_k\setminus$ curr
            \ENDIF
            \STATE curr $\gets$ curr.parent
        \ENDIF
    \ENDIF
    \ENDWHILE
    \end{algorithmic}
\end{algorithm}

\subsection{Estimating the number of image occurrences using the gap statistics principle} 
\label{Gap Statistics Section}

In many real-world scenarios, such as electron microscopy~\cite{bendory2020single,eldar2024object}, the exact number of image occurrences $K$ is unknown. 
This imposes a major challenge given that both the greedy algorithm and Algorithm~\ref{alg} require a fixed $K$. A common practice is 
solving~\eqref{eq:constrainted_optmization} for various $K$ values, choosing the one that induces the steepest change in the objective value.
This heuristic is called the ``knee'' method and has been widely adopted across diverse domains, such as clustering~\cite{tibshirani2001estimating} and regularization~\cite{hansen1992analysis}.

Here, we propose employing the gap statistics principle, which can be thought of as a statistical technique to identify the ``knee"~\cite{tibshirani2001estimating}. This principle is widely applied to a variety of domains, as discussed for example in~\cite{painsky2012exclusive,painsky2014optimal}.
The underlying idea of gap statistics is to adjust the objective value curve (as a function of the possible values of~$K$) by comparing it with its expectation under a null reference.
Let us define the gap statistic as:
\begin{equation}
    \text{gap}(K) = p(\bm{\pi}_{\text{opt}}(K)) - \mathbb{E}^{*}p(\bm{\pi}_{\text{opt}}(K)),
    \label{eq:GAP}
\end{equation}
where $p(\bm{\pi}_{\text{opt}}(K))$ represents the revenue of the optimal allocation for $K$ image occurrences (computed by Algorithm~\ref{alg}), and $\mathbb{E}^{*}p(\bm{\pi}_{\text{opt}}(K))$ is the expectation for a measurement of dimensions $N \times M$ derived from a ``null” reference distribution. To form the ``null" distribution, we randomly rearrange the pixels of the measurement, resulting in an unstructured measurement, with no template image occurrences. 
Specifically, we estimate the expectation under the null by $E^{*}p(\bm{\pi}_{\text{opt}}(K)) \approx \frac{1}{R}\sum_{r=1}^{R} p(\bm{\pi}_{\text{opt}}^{r}(K))$,
where  $p(\bm{\pi}_{\text{opt}}^{r}(K))$ is the revenue achieved by the optimal allocation, at the $r$-th rearrangement; in the experiments below, we set $R=50$. The estimate of $K$---the number of image occurrences---is the $K$ that maximizes $\text{gap}(K)$ and is denoted by $\hat{K}$.

\section{Numerical experiments}
\label{sec:Numerical_Experiments}
In this section, we conduct numerical experiments to compare Algorithm~\ref{alg} with the alternative greedy algorithm.
We use the \text{F$_{1}$}-score to evaluate the performance of both methods~\cite{ghaddar2018robust,huang2015maximum,fujino2008multi,painsky2023quality}. The $F_1$ score is defined as:
\begin{equation}
     \text{F}_1 = 2 \times \frac{\text{Precision} \times \text{TPR}}{\text{Precision} + \text{TPR}},
\end{equation}
where Precision is the ratio of correct detections divided by the total number of detections, and TPR, which stands for True Positive Rate, is the ratio of correct detections divided by the total number of image occurrences. We note that in the case where the number of image occurrences~$K$ is known, Precision equals TPR. 

We define a detection as correct if:
\begin{equation}
	\begin{aligned}
\max\left(\left|\hat{c}^{(k)}_{n} - {c}^{(k)}_{n}\right|,\left|\hat{c}^{(k)}_{m} - {c}^{(k)}_{m}\right|\right)\leq \frac{W}{2}. 
	\end{aligned}
\end{equation}
%
That is, a detection is classified as correct if the estimated image location is within $\frac{W}{2}$ pixels from the true location. This convention is common in the image detection literature~\cite{schwartzman2011multiple,cheng2017multiple}. 
We define the SNR of the measurement as
\begin{equation}
    \text{SNR} = 10\log\frac{KW^{2}}{\sigma^{2}NM},
\end{equation}
\rev{where $\sigma^2$ is the variance of the noise.}
To compare our algorithm with the greedy algorithm, which works well when the supports of the correlated image occurrences do not overlap, we define another separation condition.
\begin{equation}
	\begin{aligned}
		\label{Well Separated condition}  
		\max\left(\left| c^{(k)}_{n} - c^{(\ell)}_{n}\right|,\left|c^{(k)}_{m} - c^{(\ell)}_{m}\right|\right) \geq 2W,  \quad  \forall k \neq \ell. 
	\end{aligned}
\end{equation}
We refer to this condition as the well-separated condition.

For a given set of parameters $M,N,W,K$, we generate a measurement $y$ as follows. 
We begin by placing the first image occurrence at a random location (drawn from a uniform distribution) within the measurement.
Then, we draw another location and place the image occurrence if it meets either \eqref{eq:Dense Separation Condition} or \eqref{Well Separated condition}. This process is repeated until all $K$ images have been placed in the measurement. Finally, we add i.i.d. white Gaussian noise with zero mean and $\sigma^2$ variance to the measurement.
The code to reproduce the numerical experiments is publicly available at \href{https://github.com/saimonanuk/Optimal-detection-of-non-overlapping-images-via-combinatorial-auction}{https://github.com/saimonanuk/Optimal-detection-of-non-overlapping-images-via-combinatorial-auction}.

\subsection{A known number of image occurrences}
\begin{figure}[h]  
	\centering  \includegraphics[width=.8\columnwidth]{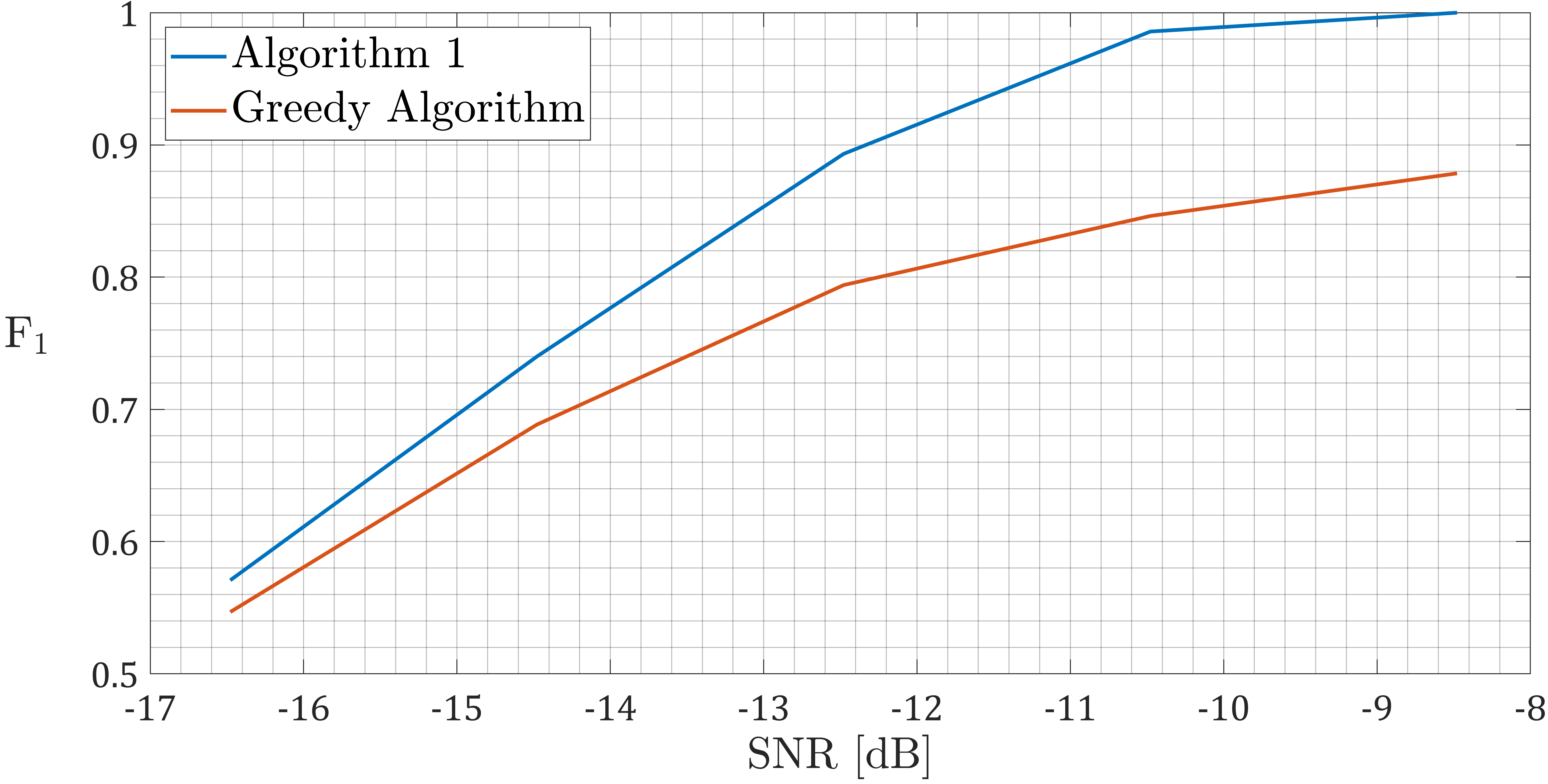}
    \caption{The average $F_{1}$ score, as a function of the SNR, of  Algorithm~\ref{alg} and the greedy algorithm, assuming the number of image occurrences is known.}
    \label{fig:F1 known particles}
\end{figure}

In the first experiment, we assume that the number of image occurrences $K$ is known, and both algorithms receive the true number of image occurrences $K$ as input. 
We set  $N=M=40$ and $K=4$, and choose a template image with all-ones entries of size $W = 3$.
We generate the measurement such that the separation condition~\eqref{eq:Dense Separation Condition} is met, and at least one pair of image occurrences is separated precisely by $W$ pixels. 
 For each  SNR level,
we conduct 1000 trials, each with a fresh measurement. 
Figure~\ref{fig:F1 known particles} shows the average $F_1$ score of Algorithm~\ref{alg} and the greedy algorithm \rev{as a function of the SNR}.
Evidently, Algorithm~\ref{alg} outperforms the greedy algorithm in all SNR regimes. As expected, in high SNR regimes, Algorithm~\ref{alg} achieves $\text{F}_1 = 1$. On the contrary, the greedy algorithm fails even in the high SNR regime, achieving $\text{F}_1 = 0.88$, due to the proximity of the image occurrences in the measurement. \rev{The average runtime per trial of Algorithm~\ref{alg} was approximately 80 seconds, whereas the average runtime of the greedy algorithm was 0.1 seconds.}

\begin{figure}[h]   
	\centering \includegraphics[width=.8\columnwidth]{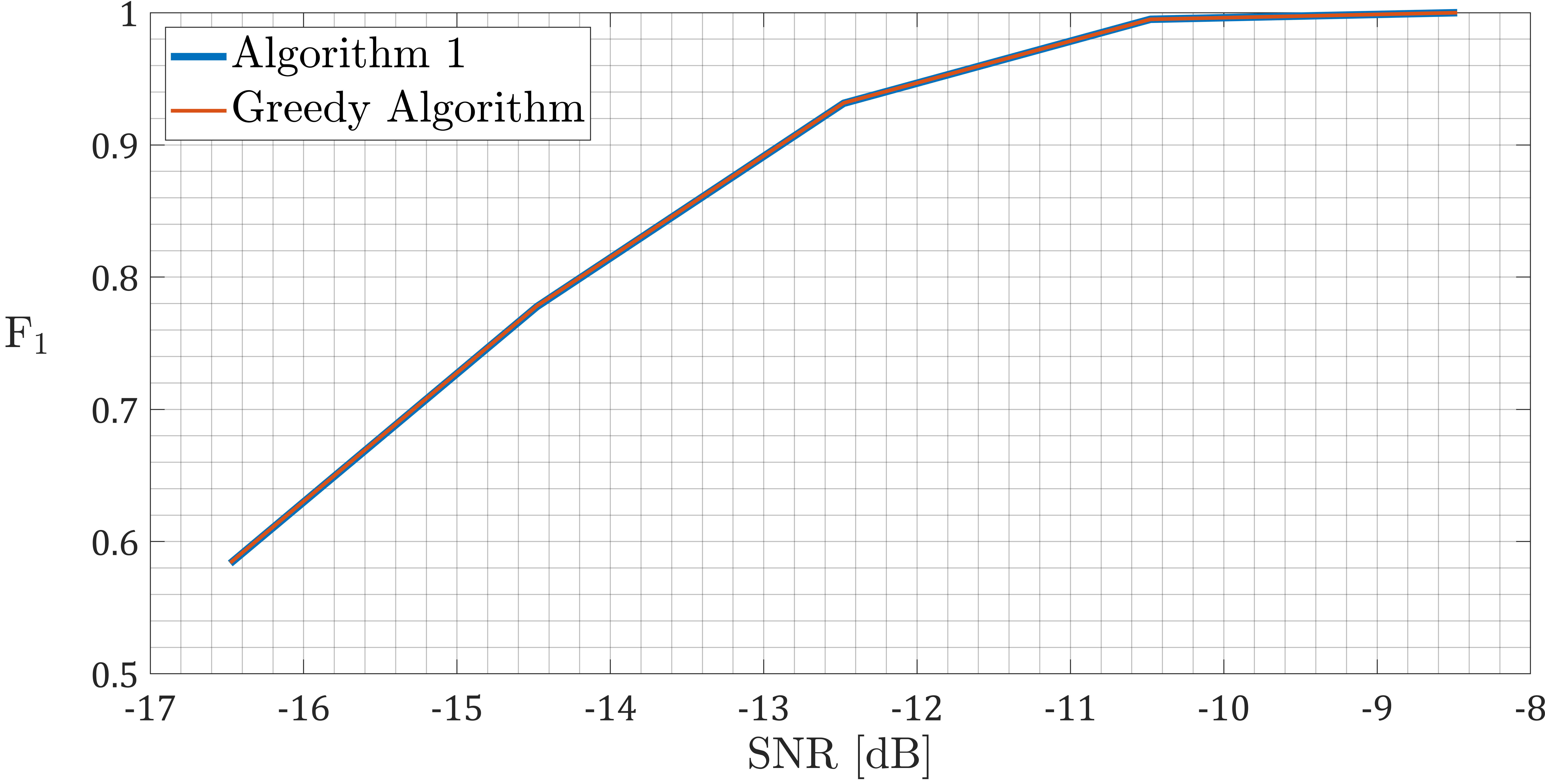}
    \caption{The average $F_{1}$ score, as a function of the SNR, of  Algorithm~\ref{alg} and the greedy algorithm, where the image occurrences satisfy the well-separated condition~\eqref{Well Separated condition}, assuming the number of image occurrences is known.}
    \label{fig:F1 known particles 2W}
\end{figure}

We repeated the same experiment, but now the image occurrences are well-separated according to~\eqref{Well Separated condition}. The results of this experiment are illustrated in Figure~\ref{fig:F1 known particles 2W}. As expected, in this case, both algorithms achieve the same results, \rev{and the running time remains the same.}

\subsection{An unknown number of image occurrences}
\begin{figure}[h]   
	\centering \includegraphics[width=.8\columnwidth]{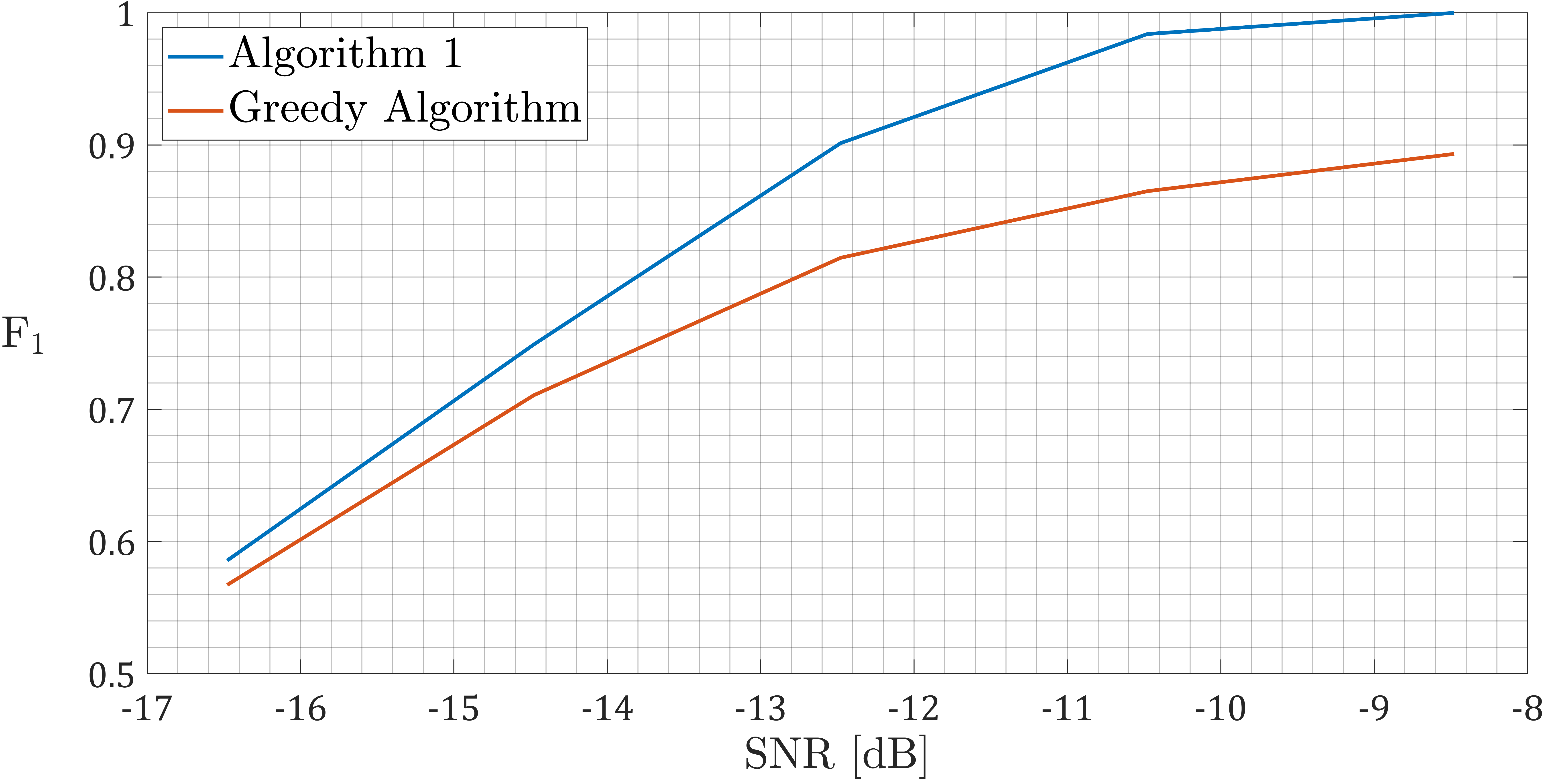}
    \caption{The average $F_{1}$ score, as a function of the SNR, of  Algorithm~\ref{alg} and the greedy algorithm assuming the number of image occurrences is unknown.}
    \label{fig:F1 unknown particles}
\end{figure}

We repeated the first experiment from the previous section (when the image occurrences satisfy~\eqref{eq:Dense Separation Condition} but not~\eqref{Well Separated condition}), but now the number of image occurrences $K$ is unknown. 
To estimate $K$, we used the gap statistics principle. This estimate is then used as an input for both algorithms to determine the image locations. The results are presented in Figure~\ref{fig:F1 unknown particles}.

\rev{Similar to} the previous case, in the unknown $K$ scenario, Algorithm~\ref{alg} outperforms the greedy algorithm in all SNR regimes. 
In addition, Algorithm~\ref{alg} provides better estimates of the number of signal occurrences $K$. 
For example, for SNR$= -12.5$, Algorithm~\ref{alg} gets $87.2\%$ accuracy in estimating $K$ precisely, while the greedy algorithm gets only $80\%$.
This gap drops with the SNR until they eventually converge to the same estimations when the SNR is low enough. 

\section{Cryo-electron microscopy examples}
\label{sec:Cryo}

\begin{figure}[h]
  \centering
  \begin{minipage}
  {0.46\columnwidth}
    \centering
    \includegraphics[width=\linewidth]{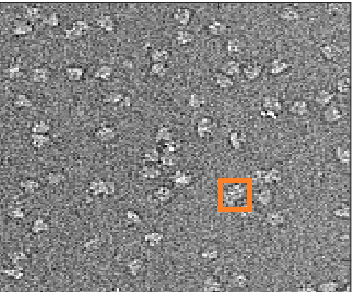}
    \label{fig:10081 dataset micrograph}
  \end{minipage}
  \hfill
  \begin{minipage}
  {0.48\columnwidth}
    \centering
    \includegraphics[width=\linewidth]{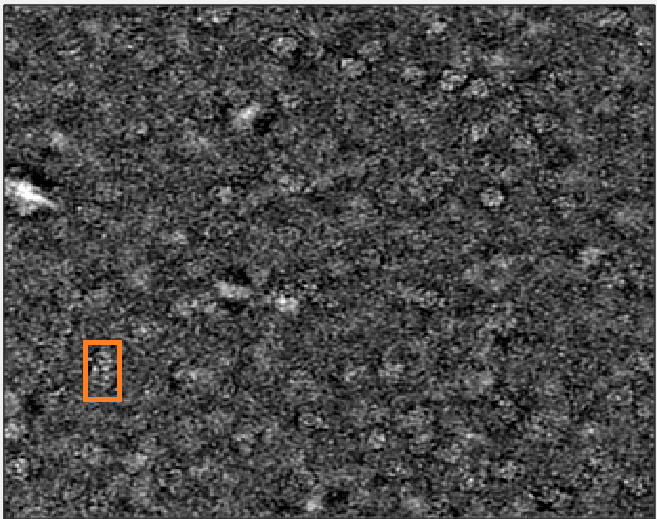}
    \label{fig:10217 dataset micrograph}
  \end{minipage}
  \caption{Two cryo-EM measurements after preprocessing steps. The left panel presents a measurement from the EMPIAR-10081 data set and the right panel shows a measurement from the EMPIAR-10217 data set. The orange squares are the dense areas used as the input for both algorithms.}
  \label{fig:cryo measurements}
\end{figure}

We applied Algorithm~\ref{alg} and the greedy algorithm to two cryo-EM datasets, available at the EMPIAR repository~\cite{iudin2016empiar}.  
We assume a disk-like shape for the images.
Following standard procedures in this field, we 
performed a sequence of preprocessing steps before applying both algorithms. 
First, we whitened the measurement by manually selecting ``noise-only" areas (i.e., without images), which were used to estimate the empirical covariance matrix of the noise. The data is then normalized by the inverse of this covariance matrix. Then, the data is downsampled by a factor of 12, concurrently reducing computational load and amplifying the SNR.
Ultimately,  we took the whitened, downsampled measurement and determined patches in which the image occurrences are not well-separated, namely, do not satisfy~\eqref{Well Separated condition}.
We focused on small dense patches since this is the scenario in which Algorithm~\ref{alg} shines. 
In well-separated \eqref{Well Separated condition} areas, it is recommended to use the computationally efficient greedy algorithm. 
Figure~\ref{fig:cryo measurements} displays two different measurements, after the preprocessing steps.

\subsection{EMPIAR 10081}
\begin{figure}[h]
  \centering
  \begin{minipage}
  {0.23\columnwidth}
    \centering
    \includegraphics[width=\linewidth]{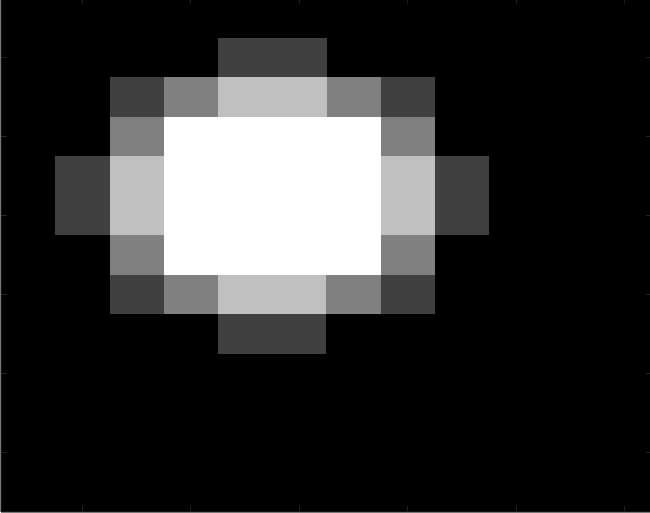}
    \label{fig:081 Greedy algorithm output}
  \end{minipage}
  \hfill
  \begin{minipage}
  {0.23\columnwidth}
    \centering
    \includegraphics[width=\linewidth]{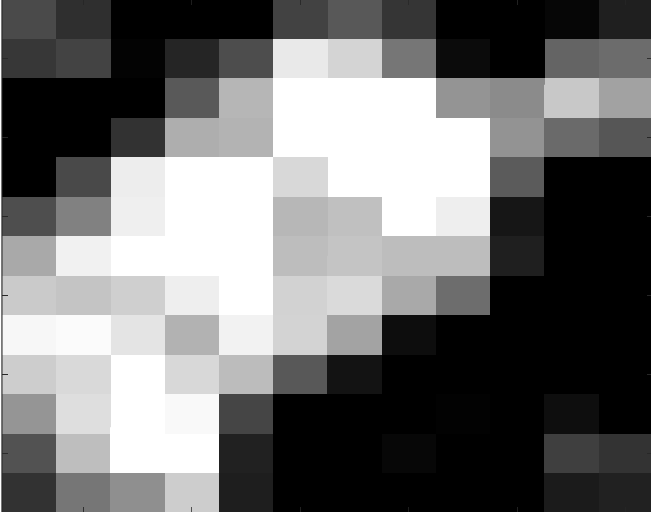}
    \label{fig:081 The original measurement}
  \end{minipage}
  \hfill
  \begin{minipage}
  {0.23\columnwidth}
    \centering
    \includegraphics[width=\linewidth]{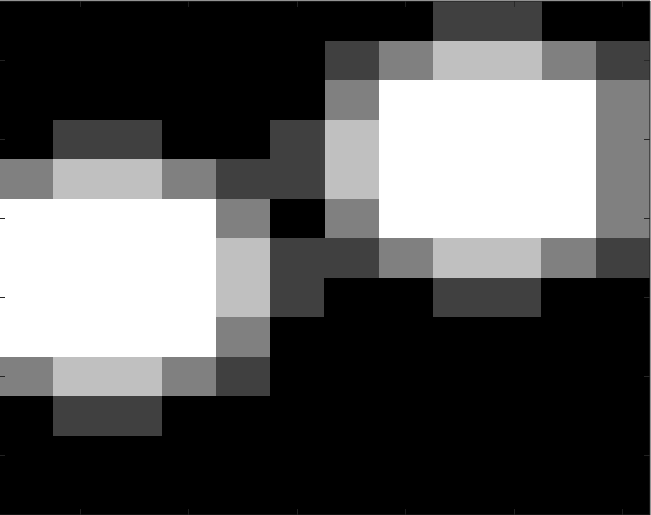}
    \label{fig:081 Algorithm 1 output}
  \end{minipage}
  \hfill
  \begin{minipage}
  {0.23\columnwidth}
    \centering
    \includegraphics[width=\linewidth]{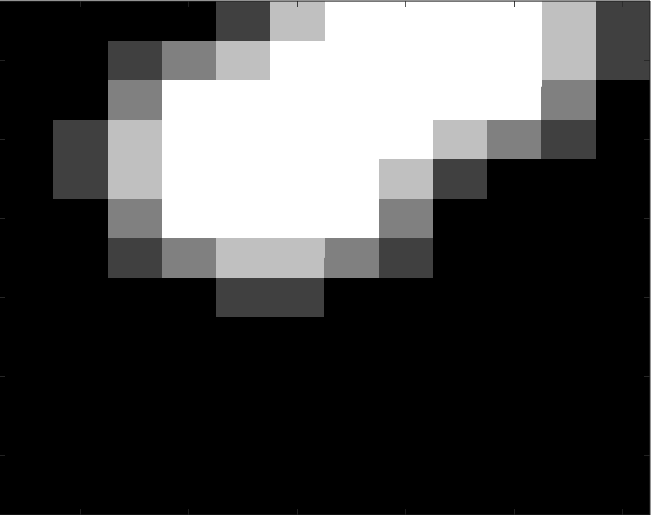}
    \label{fig:081 The original measurement}
  \end{minipage}
  \caption{\rev{From left to right: the output of the greedy algorithm, the original patch taken from the data set EMPIAR 10081 as illustrated in Figure~\ref{fig:cryo measurements}, the output of Algorithm~\ref{alg}, and the output of the cryo-EM particle picker TOPAZ~\cite{bepler2019positive}}.}
  \label{fig:081 estimation}
\end{figure}

Our first experiment is conducted on the EMPIAR-10081 data set of the Human HCN1 Hyperpolarization-Activated Channel~\cite{lee2017structures}.
The measurement is presented in Figure~\ref{fig:cryo measurements}.
We set the radius of the disc used as a template image to 3 pixels.
The number of image occurrences is estimated based on gap statistics.
Figure~\ref{fig:081 estimation} shows the output of both algorithms.  
Algorithm~\ref{alg} identifies two separated images. In contrast, the greedy algorithm identifies a single occurrence. 
\rev{For comparison, we also show the output of a popular detection algorithm for cryo-EM data sets (particle picker), called TOPAZ~\cite{bepler2019positive}. This algorithm is based on a deep learning technique (see also~\cite{dhakal2024cryotransformer}).}

\subsection{EMPIAR 10217}
\begin{figure}[h]
  \centering
  \begin{minipage}
  {0.23\columnwidth}
    \centering
    \includegraphics[width=\linewidth]{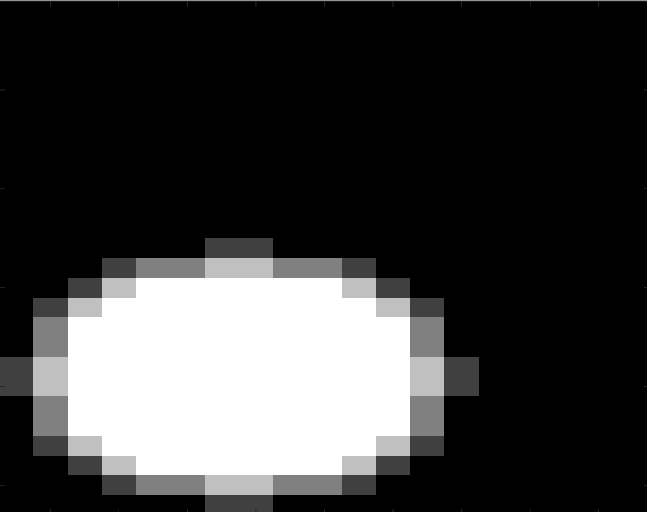}
    \label{fig:217 Greedy algorithm output}
  \end{minipage}
  \hfill
  \begin{minipage}
  {0.23\columnwidth}
    \centering
    \includegraphics[width=\linewidth]{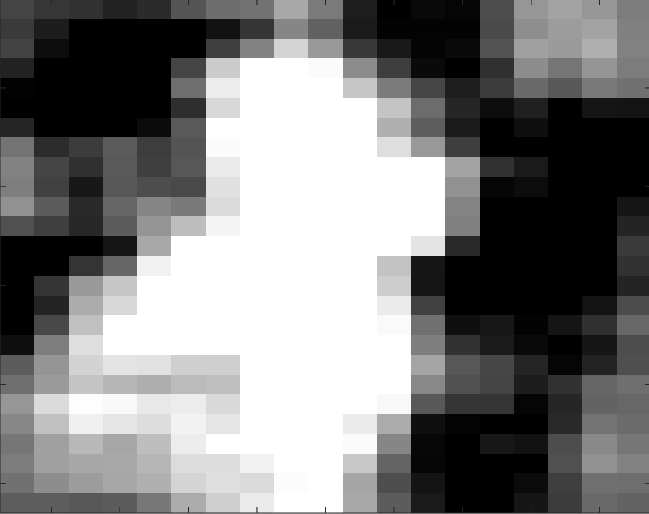}
    \label{fig:217 The original measurement}
  \end{minipage}
  \hfill
  \begin{minipage}
  {0.23\columnwidth}
    \centering
    \includegraphics[width=\linewidth]{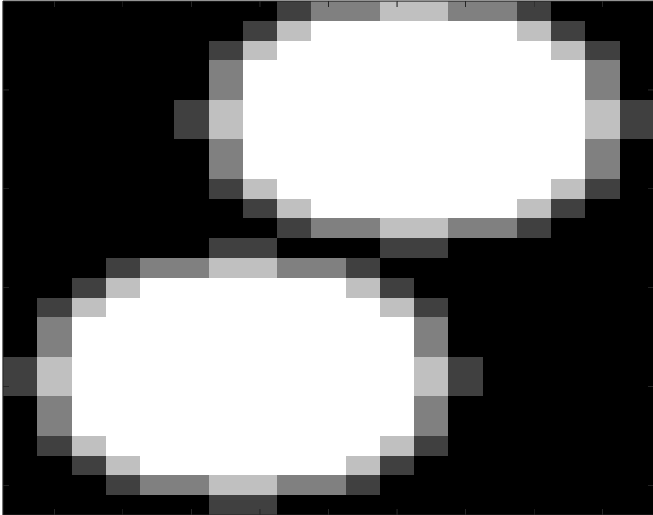}
    \label{fig:217 Algorithm 1 output.}
  \end{minipage}
  \hfill
  \begin{minipage}
  {0.23\columnwidth}
    \centering
    \includegraphics[width=\linewidth]{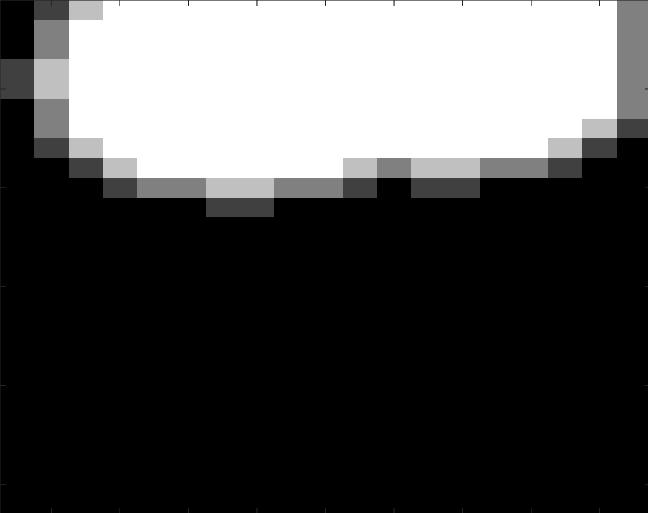}
    \label{fig:217 The original measurement}
  \end{minipage}
  \caption{\rev{From left to right: the output of the greedy algorithm, the original patch taken from the data set EMPIAR 10217 as illustrated in Figure~\ref{fig:cryo measurements}, the output of Algorithm~\ref{alg}, and the output of the cryo-EM particle picker TOPAZ~\cite{bepler2019positive}.}}
  \label{fig:217 estimation}
\end{figure}

Our second experiment is conducted on the EMPIAR-10217 data set of the bovine liver glutamate dehydrogenase~\cite{iudin2016empiar}. We set the radius of the disc used as a template image to 6 pixels.
The measurement is presented in Figure~\ref{fig:cryo measurements} and the results of both algorithms are displayed in Figure~\ref{fig:217 estimation}. 
\rev{Similar to} the previous example, Algorithm~\ref{alg} identifies two different particle images, while the greedy algorithm finds only one particle image in the measurement.

\section{CONCLUSION} \label{sec:conclusions}
This paper studies the problem of detecting multiple 
 image occurrences in a two-dimensional, noisy measurement. We approach this task by formulating it as a constrained maximum likelihood optimization problem. By showing the equivalence between this problem and a version of the winner determination problem, we design an efficient search algorithm over a binary tree. Our algorithm can find the optimal solution, even in scenarios characterized by high noise levels and densely packed image occurrences. 

Our methodology can be extended in a couple of directions that we leave for future work. First, we assume white Gaussian noise, whereas in many applications, such as cryo-EM, that noise is not white~\cite{bendory2020single}. Using the formulation of the detection problem as a combinatorial auction~\eqref{eq:WDP K bids}, changing the noise statistics will only change the revenue function~$p$, while the rest of the optimization problem remains the same. Similarly, one may want to introduce a prior on the sought locations (in the Bayesian sense). In this case, the goal will be to maximize the posterior distribution under the separation condition. This again will not change the structure of the algorithm, just the definition of the revenue. 
Second, our pruning mechanism is quite conservative since it ignores the separation condition. An important future work includes designing a more effective pruning mechanism that will accelerate the search over the tree (and thus the algorithm's runtime), perhaps at the cost of a small bounded deviation from the optimal solution.

\section*{Acknowledgments} 
T.B. is supported in part by the BSF grant no. 2020159, the NSF-BSF grant no. 2019752, and the ISF grant no. 1924/21. A.P. is supported by the in part by ISF grant no. 963/21.
The authors are grateful to Amnon Balanov for his assistance with TOPAZ. 

\bibliographystyle{plain}

\end{document}